\documentclass[10pt,technote]{IEEEtran}

\usepackage{cite,graphicx,amsmath,psfrag,bm}
\usepackage{subfigure}
\usepackage[usenames,dvipsnames]{xcolor}
\usepackage{mathabx}
\usepackage{cancel}
\usepackage{epstopdf}
\usepackage{pstool}
\usepackage{caption}
\graphicspath{{Figs/}}

\begin{document}

\title{Enhancement of Time-Reversal Subwavelength Wireless Transmission Using Pulse Shaping}

\author{Shuai Ding, Rui Zang, Lianfeng Zou, Bing-Zhong Wang,~\IEEEmembership{Member,~IEEE}
        Christophe Caloz,~\IEEEmembership{Fellow,~IEEE}
\thanks{S. Ding, L. Zou and C. Caloz are with the Department
of Electrical Engineering and Poly-Grames research center, \'{E}cole Polytechnique de Montr\'{e}al, Montr\'{e}al, Qu\'{e}bec, H3T 1J4, Canada. e-mail: shuai.ding@polymtl.ca; christophe.caloz@polymtl.ca.}
\thanks{R. Zang and B.-Z. Wang are with the Institute of Applied Physics, University
of Electronic Science and Technology of China, Chengdu 610054, China. e-mail: bzwang@uestc.edu.cn}}
\markboth{IEEE Transactions on Antennas and Propagation,~Vol.~, No.~, Month~2014}%
{Shell \MakeLowercase{\textit{et al.}}: Bare Demo of IEEEtran.cls for Journals}

\maketitle

\begin{abstract}
A novel time-reversal subwavelength transmission technique, based on pulse shaping circuits (PSCs), is proposed. Compared to previously reported approaches, this technique removes the need for complex or electrically large electromagnetic structures by generating channel diversity via pulse shaping instead of angular spectrum transformation. Moreover, the pulse shaping circuits (PSCs) are based on Radio Analog Signal Processing (R-ASP), and therefore do not suffer from the well-known issues of digital signal processing in ultrafast regimes. The proposed PSC time-reversal systems is mathematically shown to offer high channel discrimination under appropriate PSC design conditions, and is experimentally demonstrated for the case of two receivers.
\end{abstract}

\IEEEpeerreviewmaketitle
\section{Introduction}\label{Sec:Introduction}

\IEEEPARstart{T}{i}me reversal is an adaptive transmission technique with applications in many areas, such as for instance wireless communication, imaging and sensing~\cite{Bogert-Timereversal-1957,Fink-Basic-1992,Nguyen-2005-Communlett-potential}. A typical time-reversal transmission operates as follows. First, channel sounding signals are emitted from multiple sources. Next, the channel sounding signals are received by a time reversal mirror (TRM) and time reversed. Finally, the time-reversed signals are re-transmitted by the TRM. If the channel from the TRM to the sources is reciprocal, the time-reversed signals retrace the incoming path and focus at the location of the initial source.

Recent research has shown that time-reversal can lead to electromagnetic far-field sub-wavelength focusing, which represents a novel technique to overcome the Rayleigh-diffraction limit~\cite{Fink-2007-Science-Farfield}. This unique mechanism has led to interesting applications in wireless sub-wavelength transmission and far-field super-resolution detection~\cite{Ge-Super-2009,Liao-subwavelength-2009}. So far, time-reversal sub-wavelength focusing has been achieved by transforming evanescent waves into propagating waves via appropriate manipulation of the angular spectrum of the involved electromagnetic waves. This has been done by placing randomly distributed metal scatterers in the near field of the sources~\cite{Fink-2007-Science-Farfield,Ge-Super-2009,Fusco-conjugation-2010}, or electromagnetically large multilayered dielectric structures in front of the TRM~\cite{Liao-subwavelength-2009}. However, the former approach is hardly reproducible~\cite{Ge-Super-2009} while the latter requires very bulky elements.

To overcome these issues, this paper propose a novel approach to achieve time-reversal sub-wavelength transmission. This approach is based on Radio Analog Signal Processing (R-ASP), which consist in processing signals in real-time using purely analog component\cite{Caloz-2013-Analog}. R-ASP technique offers promising benefits in terms of high speed, wide bandwidth, low power consumption, high reproducibility and low cost~\cite{Caloz-2013-Analog}, an represents thus a possible alternative to digital signal processing techniques for ultrafast radio systems. The proposed approach removes the complex and large electromagnetic structure of the aforementioned time-reversal techniques while offering superior performance. This is performed using compact and efficient analog pulse shaping circuits (PSCs)~\cite{Rulikowski-2008-Arbitrary,Khaleghi-2011-Timereversalantenna} to produce channel discrimination. This paper also provides an experimental validation of the technique.

\section{Time-Reversal Transmission with Analog Signal Processing Blocks in the Receivers}\label{Sec:Timereversal}

\subsection{Time-reversal Principle, Issue, and Proposed Solution}\label{SubSec:principle}

A wireless time-reversal system consists of a time-reversal transmitter and of a number of receivers, which may be placed at subwavelength distances from each other. The operation of the system includes two successive phases: a channel calibration phase and a data transmission phase. In the channel calibration phase, all the receivers sequentially send a channel sounding pulse to the transmitter. The transmitter receives all the sounding pulses, flips them in time (time reversal), and records them. These operations constitute the \emph{channel calibration} phase, which is so called because they allow the transmitter to acquire information on the distinct channels existing between itself and the different receivers. The \emph{data transmission} phase, typically using pulse position modulation~\cite{Ge-2011-subwavelegnth}, can then start. The transmitter radiates a wave constituted of the superposition of data signals modulated onto the different sounding signals destined to the corresponding receivers, and each receiver receives its intended signal while discarding the others thanks to the sounding codes.

\begin{figure}
\begin{center}
\psfrag{T}[c][c][0.8]{Time-reversal}
\psfrag{R}[c][c][0.8]{transmitter}
\psfrag{M}[c][c][0.8]{TRM}
\psfrag{a}[c][c][0.8]{$h_{k}^\text{s}(t)$}
\psfrag{z}[c][c][0.8]{$d$}
\psfrag{A}[c][c][0.8]{Antenna-$1$}
\psfrag{B}[c][c][0.8]{Antenna-$2$}
\psfrag{D}[c][c][0.8]{Antenna-$k$}
\psfrag{E}[c][c][0.8]{Antenna-$K$-$1$}
\psfrag{F}[c][c][0.8]{Antenna-$K$}
\psfrag{p}[c][c][0.8]{PSC-$1$}
\psfrag{q}[c][c][0.8]{PSC-$2$}
\psfrag{r}[c][c][0.8]{PSC-$k$}
\psfrag{s}[c][c][0.8]{PSC-$K$-$1$}
\psfrag{t}[c][c][0.8]{PSC-$K$}
\psfrag{I}[c][c][0.8]{Receiver-$1$}
\psfrag{J}[c][c][0.8]{Receiver-$2$}
\psfrag{L}[c][c][0.8]{Receiver-$k$}
\psfrag{O}[c][c][0.8]{Receiver-$K$-$1$}
\psfrag{P}[c][c][0.8]{Receiver-$K$}
\psfrag{b}[c][c][0.7]{$h_1^\text{p}(t)$}
\psfrag{c}[c][c][0.7]{$h_{2}^\text{p}(t)$}
\psfrag{d}[c][c][0.7]{$h_{k}^\text{p}(t)$}
\psfrag{e}[c][c][0.7]{$h_{K-1}^\text{p}(t)$}
\psfrag{f}[c][c][0.7]{$h_{K}^\text{p}(t)$}
\includegraphics[width=1\columnwidth]{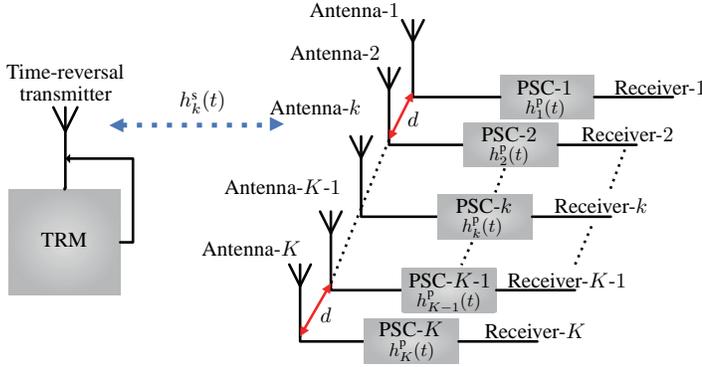}
\caption{Wireless time-reversal system.}
\label{Fig:fig1}
\end{center}
\end{figure}

If the distance between adjacent receivers is much less than the wavelength at the operation frequency, the system suffers from low spatial diversity, which leads to poor discrimination between the received signals. In order to solve this problem, we introduce here pulse shaping circuits (PSCs) with different impulse responses in each of the receivers, as shown in Fig.~\ref{Fig:fig1}. These PCSs increase the channel diversity for each receiver and hence enhance the throughput and reliability of the wireless system.

\subsection{Mathematical Description of the Proposed System}\label{SubSec:math}

\subsubsection{Channel Sounding Phase}\label{SubSec:sounding}

Receiver $k$ generates the channel sounding signal $s_k(t)$, passes it through its PSC, whose impulse response is $h_{k}^\text{p}(t)$ (`p' stands for `pulse'), and sends the resulting signal through its antenna. This radiated signal is
\begin{equation}
\label{eq:xk}
x_{k}(t)=s_k(t)\ast h_{k}^\text{p}(t),
\end{equation}
where `$\ast$' denotes the convolution product.

The time-reversal transmitter receives this signal, that has traveled across the space channel, with impulse response $h_{k}^\text{s}(t)$ (`s' stands for `space'), from receiver $k$ to itself. The received signal is then
\begin{equation}
\label{eq:yk}
y_{k}(t)=x_{k}(t)\ast h_{k}^\text{s}(t)=s(t)\ast h_{k}^\text{p}(t)\ast h_{k}^\text{s}(t),
\end{equation}
where \eqref{eq:xk} has been used in the second equality. The transmitter time-reverses the signal, which thus becomes
\begin{equation}
r_{k}(t)=y_{k}(-t)=s_{k}(-t)\ast h_{k}^\text {p}(-t)\ast h_{k}^\text {s}(-t),
\end{equation}
where \eqref{eq:yk} has been used in the second equality. This operation is sequentially repeated for all the receivers, i.e. for $k=1,2,\ldots,K$.

\subsubsection{Data Transmission Phase}\label{SubSec:data}

For each $k^\text{th}$ receiver, the transmitter pulse position modulates $r_k(t)$ by the corresponding data, which transforms $r_k(t)$ into
\begin{equation}
r_k^\text{m}(t)
=r_k(t-\tau_d)
=r_k(t)\ast\delta(t-\tau_d)
\end{equation}
(`m' stands for `modulation'), where $\tau_d$, $d=1,2,\ldots,D$ is the delay associated with the corresponding $D$-ary symbol of the data stream. The transmitter then sends the modulated signals through its antenna. All the signals, $k=1,2,\ldots,K$, can be transmitted \emph{simultaneously}, as they will be discriminated by the sounding codes, as shall be seen next.

Denoting the channel from the transmitter to the end of receiver $k$, $h_k(t)$, one has
\begin{equation}
h_k(t)=h_k^\text{s}(t)\ast h_k^\text{p}(t).
\end{equation}
The signal received by the $k^\text{th}$ receiver from the modulated  signal $r_k^\text{m}(t)$, which is destined to it, is then
\begin{equation}
\label{eq:fkt}
\begin{split}
  f_{k}(t)&= r_{k}^\text{m}(t)\ast h_{k}(t)\\
  &=\{[s_k(-t)\ast h_{k}^\text {p}(-t)\ast h_{k}^\text{s}(-t)]\ast\delta(t-\tau_{kd})\}\\
  &\qquad\qquad\ast [h_{k}^\text{s}(t)\ast h_{k}^\text{p}(t)],
\end{split}
\end{equation}
while the signal received by the $l^\text{th}$ ($l\neq k$) receiver from the modulated signal $r_k^\text{m}(t)$, which is not destined to it, is
\begin{equation}
\label{eq:fkl}
\begin{split}
  f_{l}(t)&= r_{k}^\text{m}(t)\ast h_{l}(t)\\
  &=\{[s_k(-t)\ast h_{k}^\text {p}(-t)\ast h_{k}^\text{s}(-t)]\ast\delta(t-\tau_{kd})\}\\
  &\qquad\qquad\ast [h_{l}^\text{s}(t)\ast h_{l}^\text{p}(t)].
\end{split}
\end{equation}

Using the commutativity and associativity properties of the convolution product, expressions~\eqref{eq:fkt} and~\eqref{eq:fkl} may rewritten as
\begin{subequations}\label{eq:fklt_reform}
\begin{equation}
f_{k}(t)
=s_k(-t)\ast[h_k^\text{s}(-t)\ast h_{k}^\text{s}(t)\ast\delta(t-\tau_{kd})]\ast h_{k}^\text{p}(-t)\ast h_{k}^\text{p}(t),
\end{equation}
and
\begin{equation}
f_{l}(t)
=s_k(-t)\ast h_k^\text{p}(-t)\ast[h_{k}^\text{s}(-t)\ast h_{l}^\text{s}(t)\ast\delta(t-\tau_{kd})]\ast h_{l}^\text{p}(t).
\end{equation}
\end{subequations}
Since the the spatial channel responses $h_{k}^\text{s}(t)$ and $h_{l}^\text{s}(t)$ are highly correlated for a sub-wavelength array \cite{Ding-2013-subwavelegnth}, we can make the following approximation:
\begin{equation}
h_{k}^\text{s}(-t)\ast h_{k}^\text{s}(t)\ast\delta(t-\tau_{kd})\approx h_{k}^\text{s}(-t)\ast h_{l}^\text{s}(t)\ast\delta(t-\tau_{kd})=H(t).
\end{equation}
With this definition of $H(t)$, Eqs.~\eqref{eq:fklt_reform} can be re-written as
\begin{subequations}\label{eq:fkl_fin}
\begin{equation}\label{eq:fk_fin}
f_{k}(t)=z(t)\ast h_{k}^\text{p}(-t)\ast h_{k}^\text{p}(t),
\end{equation}
and
\begin{equation}\label{eq:fl_fin}
f_{l}(t)=z(t)\ast h_{k}^\text{p}(-t)\ast h_{l}^\text{p}(t),
\end{equation}
where
\begin{equation}
z(t)=s_k(-t)\ast H(t).
\end{equation}
\end{subequations}
Equations~\eqref{eq:fk_fin} and~\eqref{eq:fl_fin} reveal that the level of discrimination between $f_{k}(t)$ and $f_{l}(t)$ depends on the degree of dissimilarity between $h_{k}^\text{p}(t)$ and $h_{l}^\text{p}(t)$. Maximizing discrimination implies minimizing the correlation between $h_{k}^\text{p}(t)$ and $h_{l}^\text{p}(t)$.

Low correlated responses $h_{k}^\text{p}(t)$ and $h_{l}^\text{p}(t)$ may be produced by digital signal processing (DSP) or analog signal processing (ASP) PSCs. However, DSP suffers from limited speed and high power consumption due to analog-digital/digital-analog converters. Therefore, we choose here R-ASP technology~\cite{Caloz-2013-Analog,Shulabh-Chipless-2011,Gupta-CSection-MT} to build the PSCs of the receivers.

\section{Channel Discrimination Enhancement \\ using Pulse Shaping}\label{Sec:Resolution}

\begin{figure}
\begin{center}
\psfrag{a}[c][c][0.85]{Input}
\psfrag{b}[c][c][0.85]{~~~Output}
\psfrag{c}[c][c][0.85]{Power}
\psfrag{d}[c][c][0.85]{splitter}
\psfrag{e}[c][c][0.85]{Power}
\psfrag{f}[c][c][0.85]{combiner}
\psfrag{g}[c][c][0.85]{$a_{k1}, \Delta\tau_{k1}$}
\psfrag{h}[c][c][0.85]{$a_{k2}, \Delta\tau_{k2}$}
\psfrag{i}[c][c][0.85]{$a_{k3}, \Delta\tau_{ki}$}
\psfrag{j}[c][c][0.85]{$~a_{k(M-1)}, \Delta\tau_{k(M-1)}$}
\psfrag{k}[c][c][0.85]{$~a_{kM}, \Delta\tau_{kM}$}
\psfrag{M}[c][c][0.85]{Delay block}
\includegraphics[width=1\columnwidth]{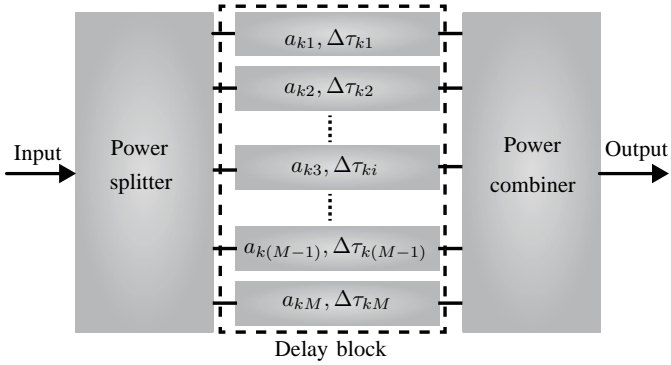}
\caption{$k^\text{th}$ pulse shaping circuit (PSC).}
\label{Fig:fig2}
\end{center}
\end{figure}

Figure~\ref{Fig:fig2} shows the block diagram of the PSC of the $k^\text{th}$ receiver. The PSC consists of three blocks: 1)~an input 1:$M$ power splitter, which divides the input pulse into $M$ identical pulses; 2)~a delay block, consisting of $M$ passive non-dispersive delay lines which each delay the $i^\text{th}$ pulse by $\Delta\tau_{ki}$ ($i=1,2,\ldots,M$); 3)~an output $M$:1 power combiner, which adds up the delayed pulses into a single signal. Since the delay block is non-dispersive, the transfer function of the $k^\text{th}$ PSC is simply
\begin{equation}
\label{eq:Hk}
H_{k}^\text{p}(\omega)
=\sum_{i=1}^{M}a_{ki}e^{-j\omega\Delta\tau_{ki}},
\end{equation}
where $a_{ki}$ is the magnitude coefficient of the $i^\text{th}$ channel. Here, all the magnitude coefficients are assumed to be equal, i.e.
\begin{equation}\label{eq:aki}
a_{k1}=a_{k2}=\cdots=a_{ki}=\cdots=a_{kM-1}=a_{kM}=a.
\end{equation}
Since in addition the input pulse is equally split in $M$ pulses, the impulse response of the PSC is readily obtained as the inverse Fourier transform of~\eqref{eq:Hk} as
\begin{equation}\label{eq:hkpt}
h_{k}^\text{p}(t)=\sum_{i=1}^{M}a\delta(t-\Delta\tau_{ki}).
\end{equation}
In order to facilitate the forthcoming analysis, we arrange the delay lines such that
\begin{equation}\label{eq:tau}
\Delta\tau_{k1}\leq\Delta\tau_{k2}\leq\cdots\leq\Delta\tau_{ki}\leq\Delta\tau_{kM-1}\leq\Delta\tau_{kM}.
\end{equation}
As mentioned in Sec.~\ref{Sec:Timereversal}, the level of discrimination between $f_{k}(t)$ and $f_{l}(t)$ depends on the degree of dissimilarity between $h_{k}^\text{p}(t)$ and $h_{l}^\text{p}(t)$. According to~\eqref{eq:fkl_fin}, $f_{k}(t)$ and $f_{l}(t)$ differ only by their respective terms $h_{k}^\text{p}(-t)\ast h_{k}^\text{p}(t)$ and $h_{k}^\text{p}(-t)\ast h_{l}^\text{p}(t)$, and these terms represent therefore the functions to be compared in order to find the conditions for maximal channel discrimination. These two terms are in fact the auto-correlation and cross-correlation functions of $h_{k}^\text{p}(t)$ and between $h_{k}^\text{p}(t)$ and $h_{l}^\text{p}(t)$, respectively, which may be noted
\begin{subequations}\label{eq:corr}
\begin{equation}\label{eq:corrhkk}
c_{k,k}(t)=h_{k}^\text{p}(-t)\ast h_{k}^\text{p}(t)
\end{equation}
and
\begin{equation}\label{eq:corrhkl}
c_{k,l}(t)=h_{k}^\text{p}(-t)\ast h_{l}^\text{p}(t).
\end{equation}
\end{subequations}
Substituting~\eqref{eq:hkpt} into~\eqref{eq:corr} and using the even property of the Dirac delta function yields
\begin{subequations}\label{eq:exc}
\begin{equation}\label{eq:exckk}
\begin{split}
c_{k,k}(t)&=\sum_{i=1}^M\sum_{j=1}^{M}a^{2}\delta(t+\Delta\tau_{ki})\ast\delta(t-\Delta\tau_{kj})\\
&=\sum_{i=1}^M\sum_{j=1}^{M}a^{2}\delta(t-\Delta\tau_{kj}+\Delta\tau_{ki})\\
&=Ma^{2}\delta(t)+\sum_{i=1}^M\sum_{j=1,j\neq i}^{M}a^{2}\delta(t-\Delta\tau_{kj}+\Delta\tau_{ki})
\end{split}
\end{equation}
and
\begin{equation}\label{eq:exckl}
\begin{split}
c_{k,l}(t)&=\sum_{i=1}^M\sum_{j=1}^{M}a^{2}\delta(t+\Delta\tau_{ki})\ast\delta(t-\Delta\tau_{lj})\\
&=\sum_{i=1}^M\sum_{j=1}^{M}a^{2}\delta(t-\Delta\tau_{lj}+\Delta\tau_{ki})\\
&=\sum_{i=1}^{M}a^{2}\delta(t-\Delta\tau_{li}+\Delta\tau_{ki})\\
&\qquad+\sum_{i=1}^M\sum_{j=1,j\neq i}^{M}a^{2}\delta(t-\Delta\tau_{lj}+\Delta\tau_{ki}).
\end{split}
\end{equation}
\end{subequations}
By definition of the auto-correlation, $c_{k,k}(t)$ reaches its peak (maximum) at $t=0$.

Consider $c_{k,l}(t)$. If
\begin{equation}\label{eq:taun}
\Delta\tau_{l1}-\Delta\tau_{k1}=\cdots=\Delta\tau_{li}-\Delta\tau_{ki}=\cdots=\Delta\tau_{lM}-\Delta\tau_{kM}=C_{kl},
\end{equation}
where $C_{kl}$ is independent of $i,j$, $\forall k,l$, then
\begin{equation}\label{eq:cklnc}
c_{k,l}(t)=Ma^{2}\delta(t-C_{kl})+\sum_{i=1}^M\sum_{j=1,j\neq i}^{M}a^{2}\delta(t-\Delta\tau_{kj}+\Delta\tau_{ki}).
\end{equation}
In this case, comparison with~\eqref{eq:exckk} indicates that the peak of $c_{k,l}(t)$ will be as high as the peak of $c_{k,k}(t)$\footnote{Only the first term in~{eq:cklnc} and ~\eqref{eq:exckk} contribute to the peak of the function since in the second terms the Dirac delta function terms lead to maxima that are distributed in time.}, which means that $h_{k}^\text{p}(t)$ and $h_{l}^\text{p}(t)$ are fully correlated.

If, in contrast,
\begin{equation}\label{eq:taun}
\Delta\tau_{l1}-\Delta\tau_{k1}\neq\cdots\neq\Delta\tau_{li}-\Delta\tau_{ki}\neq\cdots\neq\Delta\tau_{lM}-\Delta\tau_{kM},
\end{equation}
the maxima in the first term of~\eqref{eq:exckl} are completely spread out in time (all the terms contribute at different times), then this first term is much smaller than the first term in~\eqref{eq:exckk} and therefore $h_{k}^\text{p}(t)$ and $h_{l}^\text{p}(t)$ are minimally correlated.

Thus, maximal channel discrimination is achieved by designing the PSC such that condition~\eqref{eq:taun} is satisfied.

\section{System Benefits and Features}\label{sec:benefits}

The proposed PSC time reversal enhancement technique offers several compared to other techniques:
\begin{enumerate}
  \item Compared with DSP techniques, its R-ASP nature provides larger bandwidth, faster response and lower cost.

  \item In contrast to other time-reversal approaches, it does not require an electrically large time-reversal transmitter (also called TRM), since channel diversity is provided by pulse shaping instead of TRM aperture.

  \item In contrast to other time-reversal approaches, it does not require complex electromagnetic structures (e.g. metallic ``brush-like'' structure in~\cite{Fink-2007-Science-Farfield} or multilayer radome~\cite{Liao-subwavelength-2009}), but simple and integrable components (the PSCs) which in addition provide large degrees of flexibilities to handle large numbers of channels.
\end{enumerate}

\section{Experiment Demonstration}\label{Sec:experiment}

\begin{figure}
\begin{center}
\subfigure[]{\label{fig:loop1}
\psfrag{b}[l][c][0.75]{Time-reversal}
\psfrag{a}[l][c][0.75]{transmitter}
\psfrag{c}[c][c][0.75]{Antenna-$1$}
\psfrag{d}[c][c][0.75]{Antenna-$2$}
\psfrag{s}[l][c][0.75]{$d\leq \lambda_0$}
\psfrag{e}[l][c][0.75]{Receiving antennas}
\psfrag{i}[c][c][0.75]{PSC-$1$}
\psfrag{f}[c][c][0.75]{PSC-$2$}
\psfrag{g}[c][c][0.75]{Filter}
\psfrag{h}[c][c][0.75]{Filter}
\psfrag{N}[l][c][0.7]{LNA}
\psfrag{L}[l][c][0.7]{LNA}
\psfrag{D}[c][c][0.75]{DSA}
\psfrag{S}[c][c][0.75]{72004B}
\psfrag{A}[c][c][0.75]{AWG}
\psfrag{W}[c][c][0.75]{7122B}
\psfrag{C}[c][c][0.75]{Computer}
\psfrag{I}[c][c][0.75]{Indoor environment}
\psfrag{P}[c][c][0.75]{PA}
\includegraphics[width=1\columnwidth]{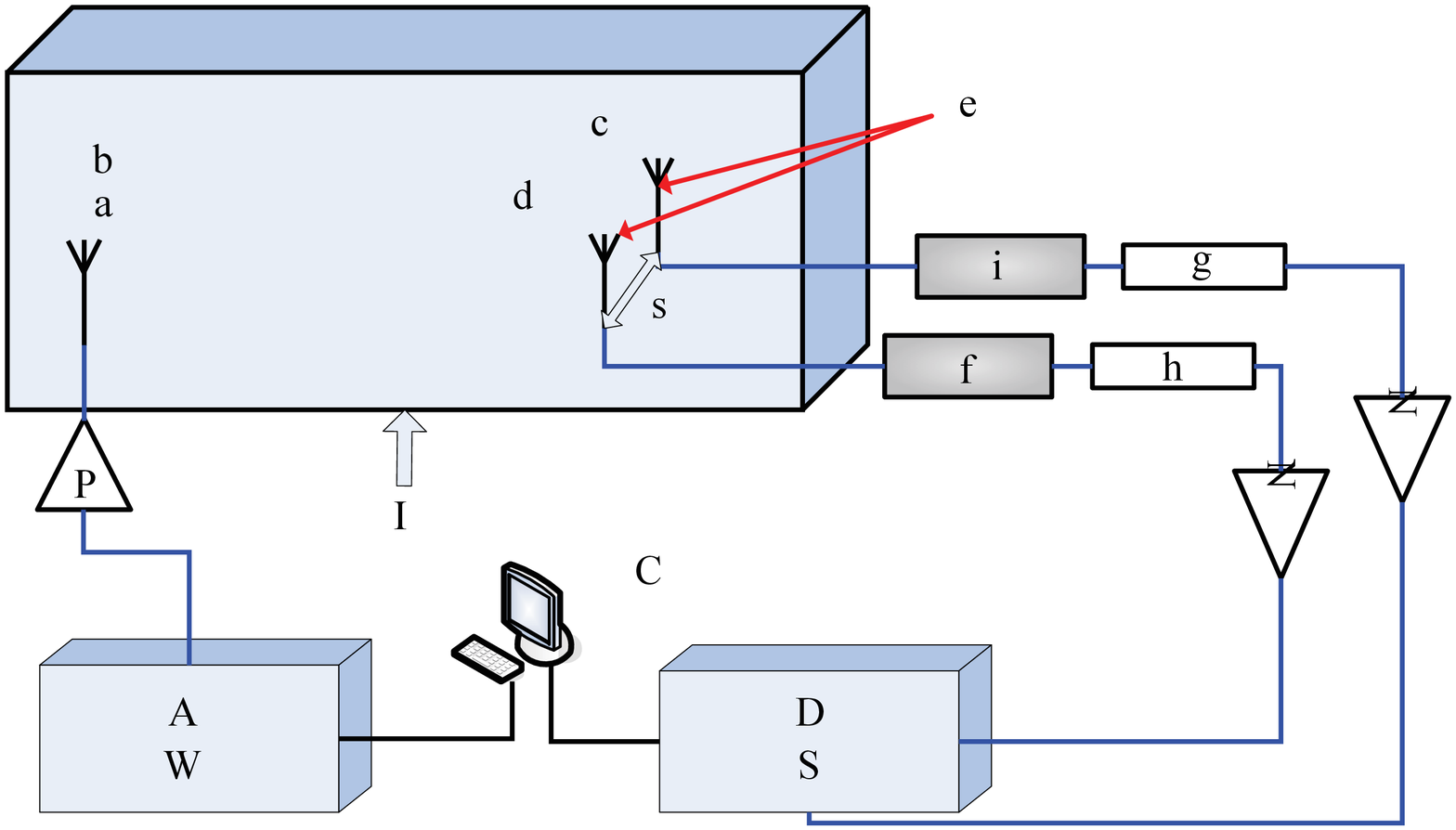}}
\subfigure[]{\label{fig:loop2}
\psfrag{a}[l][c][1]{\textcolor[rgb]{1.00,0.00,0.00}{$d\ll \lambda_0$}}
\psfrag{b}[l][c][1]{Antenna-$1$}
\psfrag{c}[l][c][1]{Antenna-$2$}
\includegraphics[width=0.7\columnwidth]{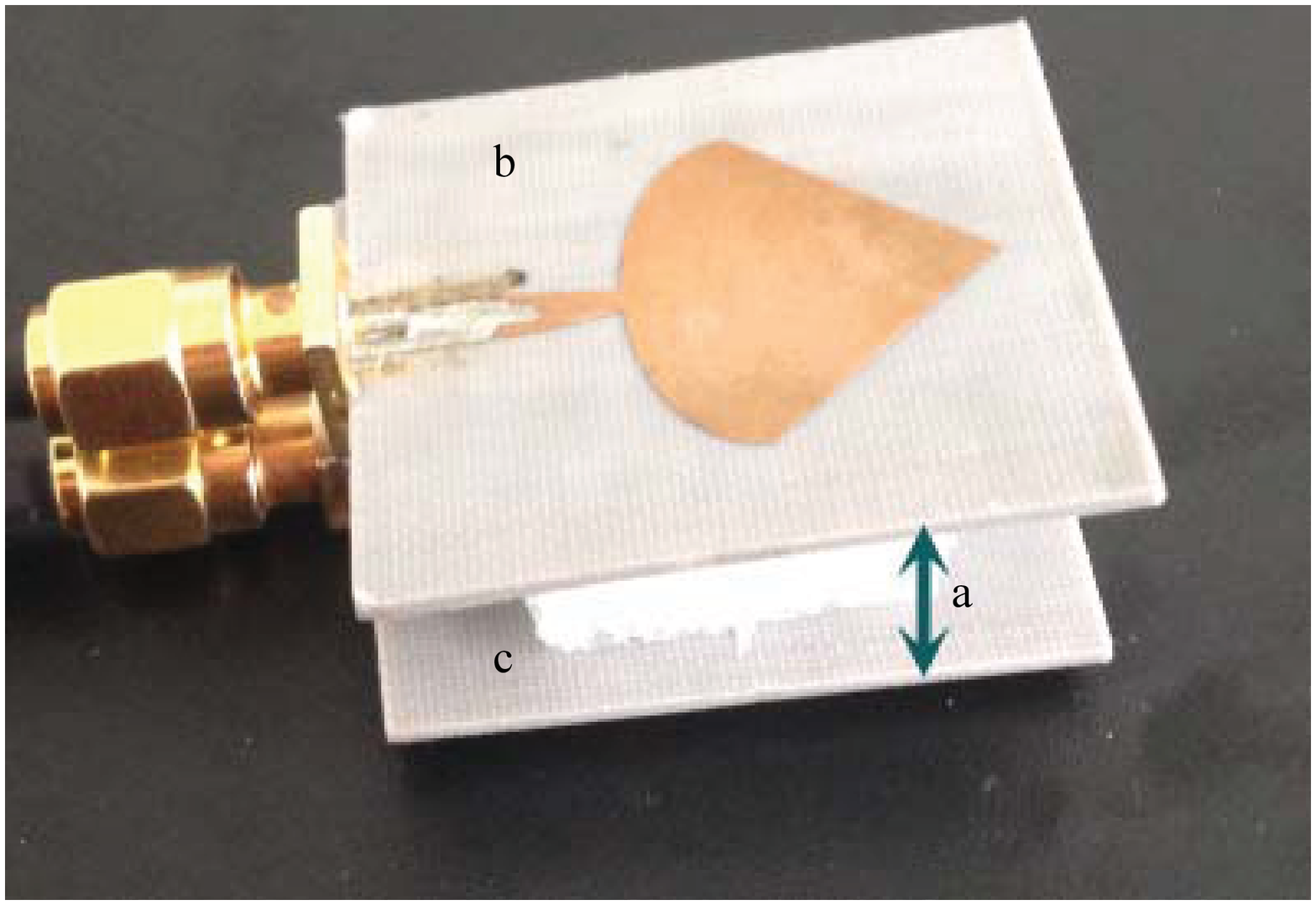}}
\subfigure[]{\label{fig:loop3}
\psfrag{a}[c][c][0.8]{Input block}
\psfrag{b}[c][c][0.8]{Output block}
\psfrag{c}[c][c][0.8]{Delay block}
\includegraphics[width=0.8\columnwidth]{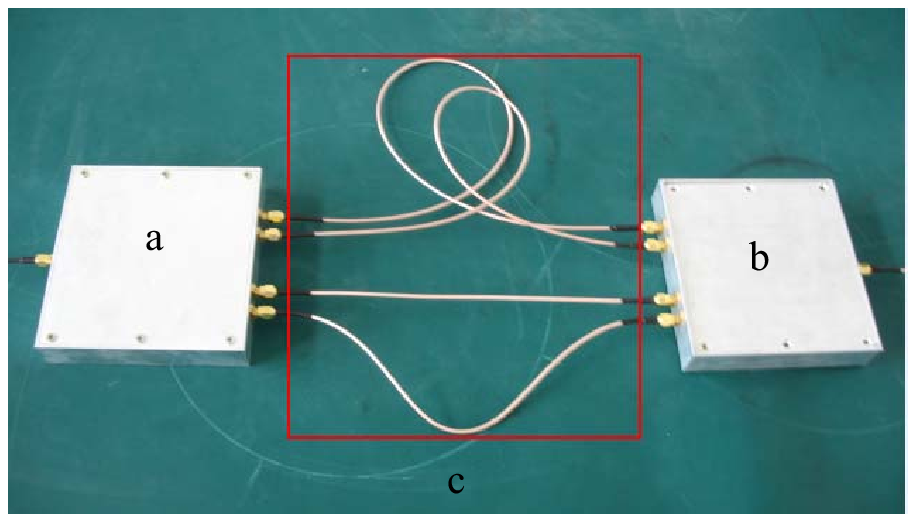}}
\caption{Experiment. a) Schematic of the setup. b) Receiving antennas. c) One of the two pulse shaping circuits (PSCs).}\label{Fig:setup}
\end{center}
\end{figure}

Figure~\ref{fig:loop1} shows the schematic experimental setup, corresponding to~Fig.~\ref{Fig:fig1} for the case $K=2$. The experiments are carried out in an indoor environment. A modulated Gaussian pulse with $5$~GHz central frequency and $3$~GHz bandwidth is selected as the channel sounding pulse. The time-reversal transmitter uses an arbitrary waveform generator (Tektronix AWG7122B) to emulate the time-reversed sounding pulses and a $3.1-10.6$~GHz UWB antenna for transmitting them. The two receivers receive the signal using UWB antennas, shown in Fig.~\ref{fig:loop2} (similar to that of the transmitter), incorporate the PSCs shown in Fig.~\ref{fig:loop3} for pulse shaping, and are connected to a digital serial analyzer (Tektronix DSA72004B) for data analysis. The distance between the transmit and receive antennas is $10$~mm, which is far less than the wavelength of the operation: $f=6.5$ to $3.5$~GHz $\rightarrow$ $\lambda_0=46$ to $86$~mm  $\rightarrow$ $d=\lambda_0/46$ to $\lambda_0/86$.
The PSCs [Fig.~\ref{fig:loop3}] are built as follows. An input $1:4$ power splitter (DC to $8$~GHz operation frequency range) splits the input pulse into four identical pulses. The same device is used as the output $4$:$1$ power combiner to sum up the delayed pulses. The delay block consists of $4$ delay lines of different lengths selected so as to maximize channel discrimination according to~\eqref{eq:taun}. The corresponding parameters are listed in Tab.~\ref{Tab:Para}.
\begin{table}[t!]
\caption{Parameters of the PSCs}
\label{Tab:Para}
\centering
\begin{tabular}{c|c}
\hline
\bfseries PSC-$1$ & \bfseries PSC-$2$\\
\hline
Delay $\Delta \tau_{\text{1}i}$  & Delay $\Delta \tau_{\text{2}i}$ \\
\hline
0.9 ns  &  0.3 ns\\
\hline
0.9 ns  &  0.6 ns\\
\hline
0.9 ns  &  0.9 ns\\
\hline
0.9 ns  &  1.2 ns\\
\hline
\end{tabular}
\end{table}

\begin{figure}
\begin{center}
\subfigure[]{\label{fig:a11}
\psfrag{A}[l][l][0.85]{Antenna-$1$}
\includegraphics[width=0.48\columnwidth]{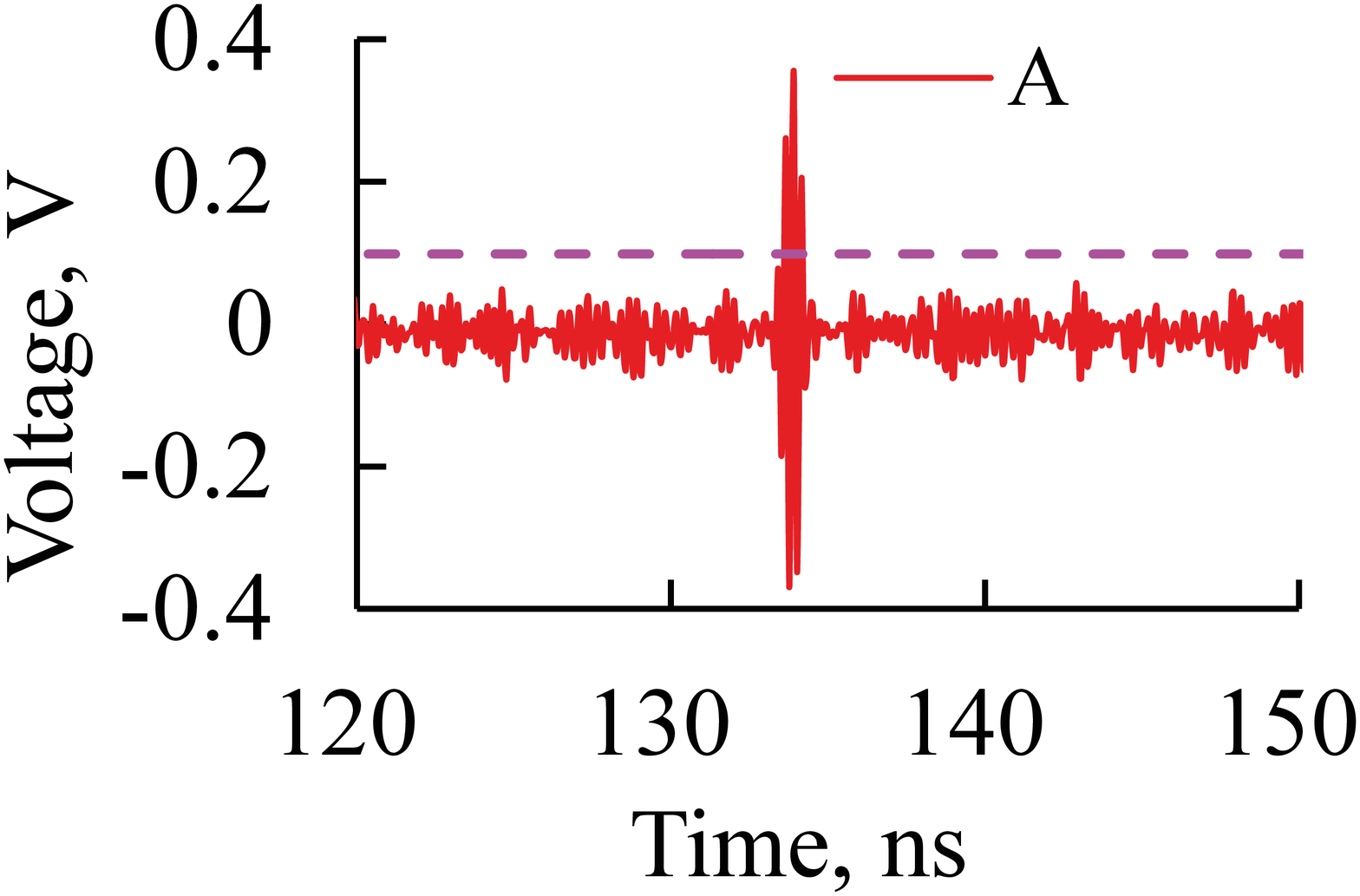}}
\subfigure[]{\label{fig:a21}
\psfrag{A}[l][l][0.85]{Antenna-$2$}
\includegraphics[width=0.48\columnwidth]{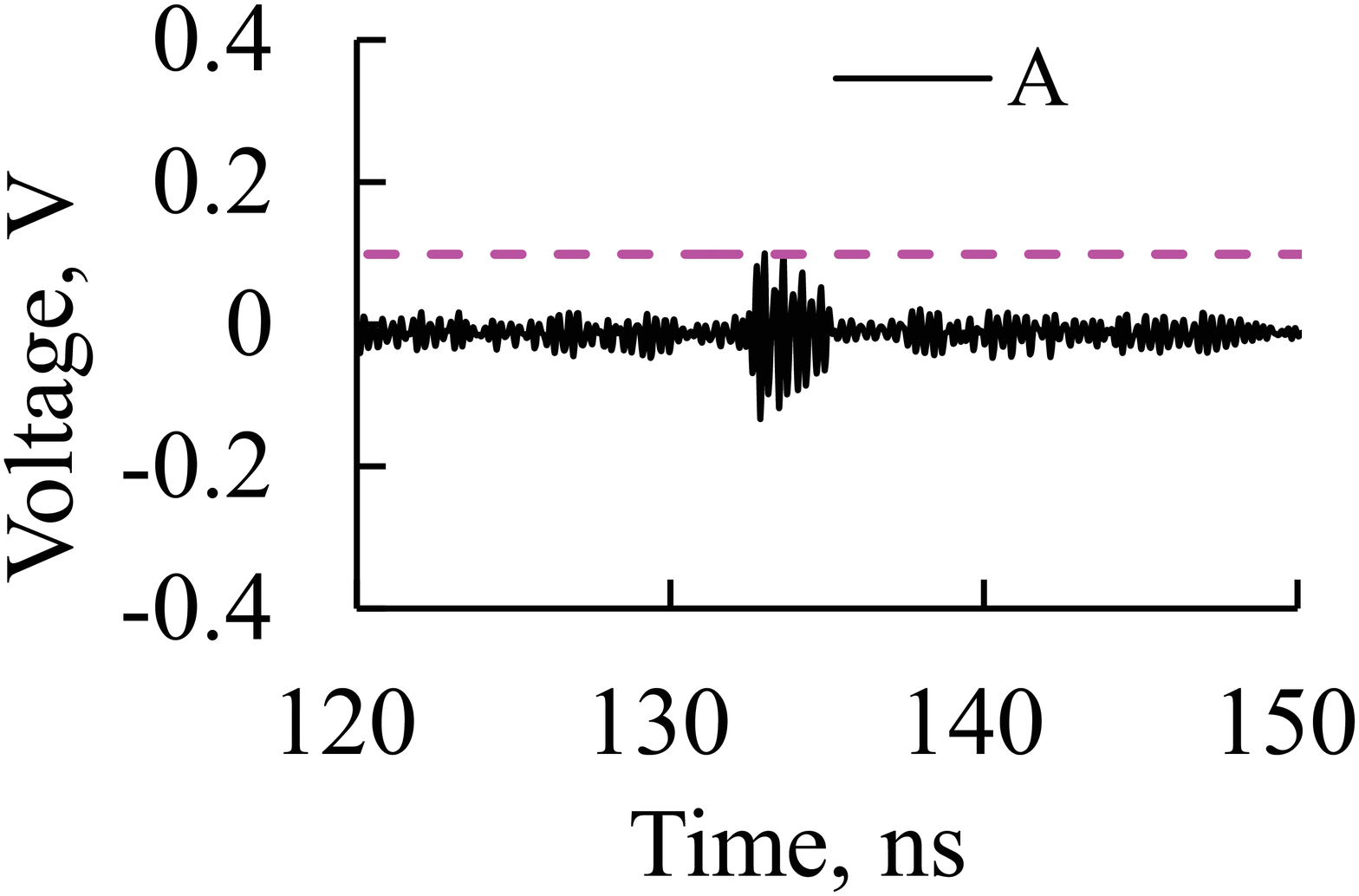}}
\subfigure[]{\label{fig:a11}
\psfrag{A}[l][l][0.85]{Antenna-$1$}
\includegraphics[width=0.48\columnwidth]{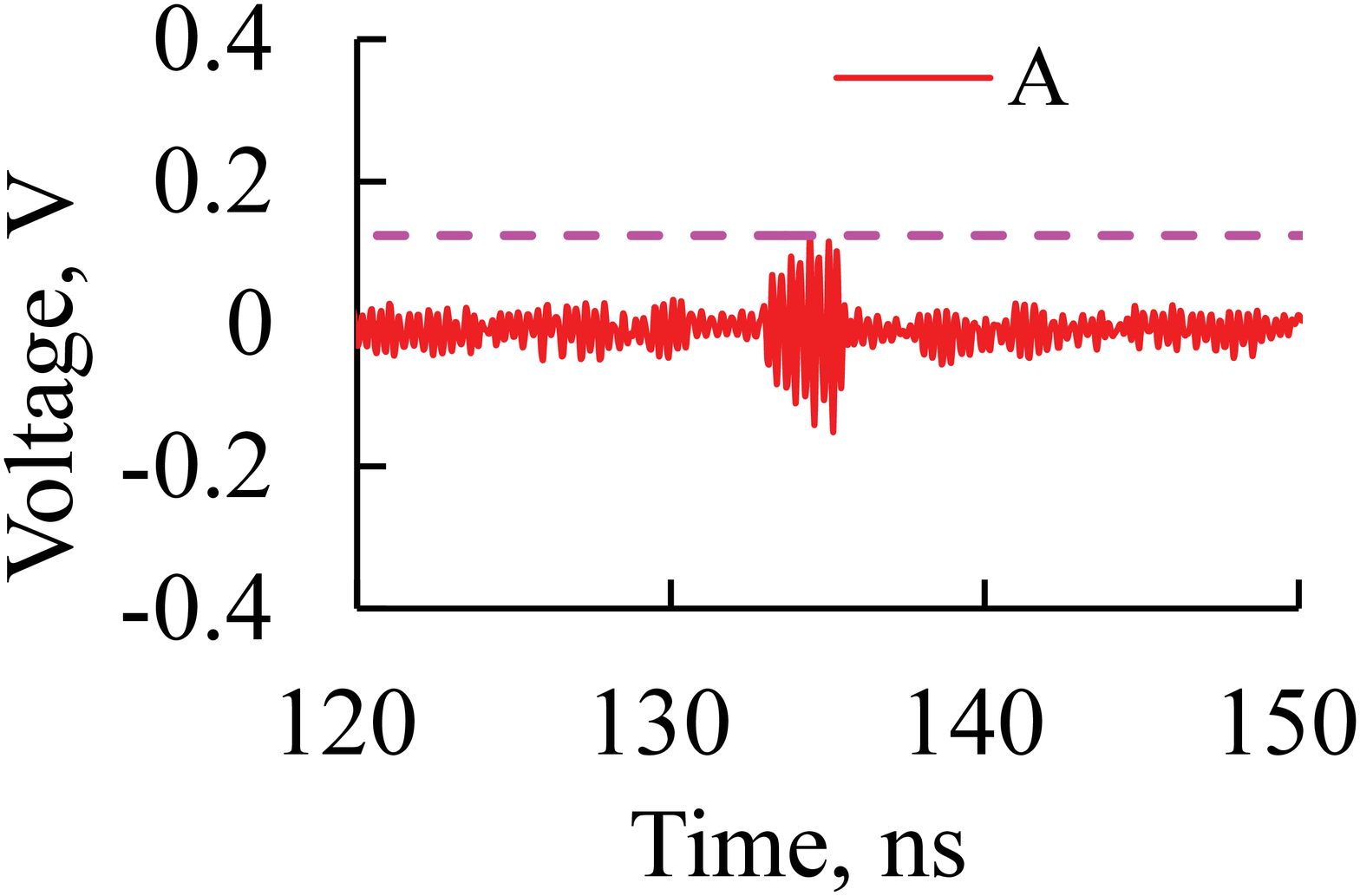}}
\subfigure[]{\label{fig:a11}
\psfrag{A}[l][l][0.85]{Antenna-$2$}
\includegraphics[width=0.48\columnwidth]{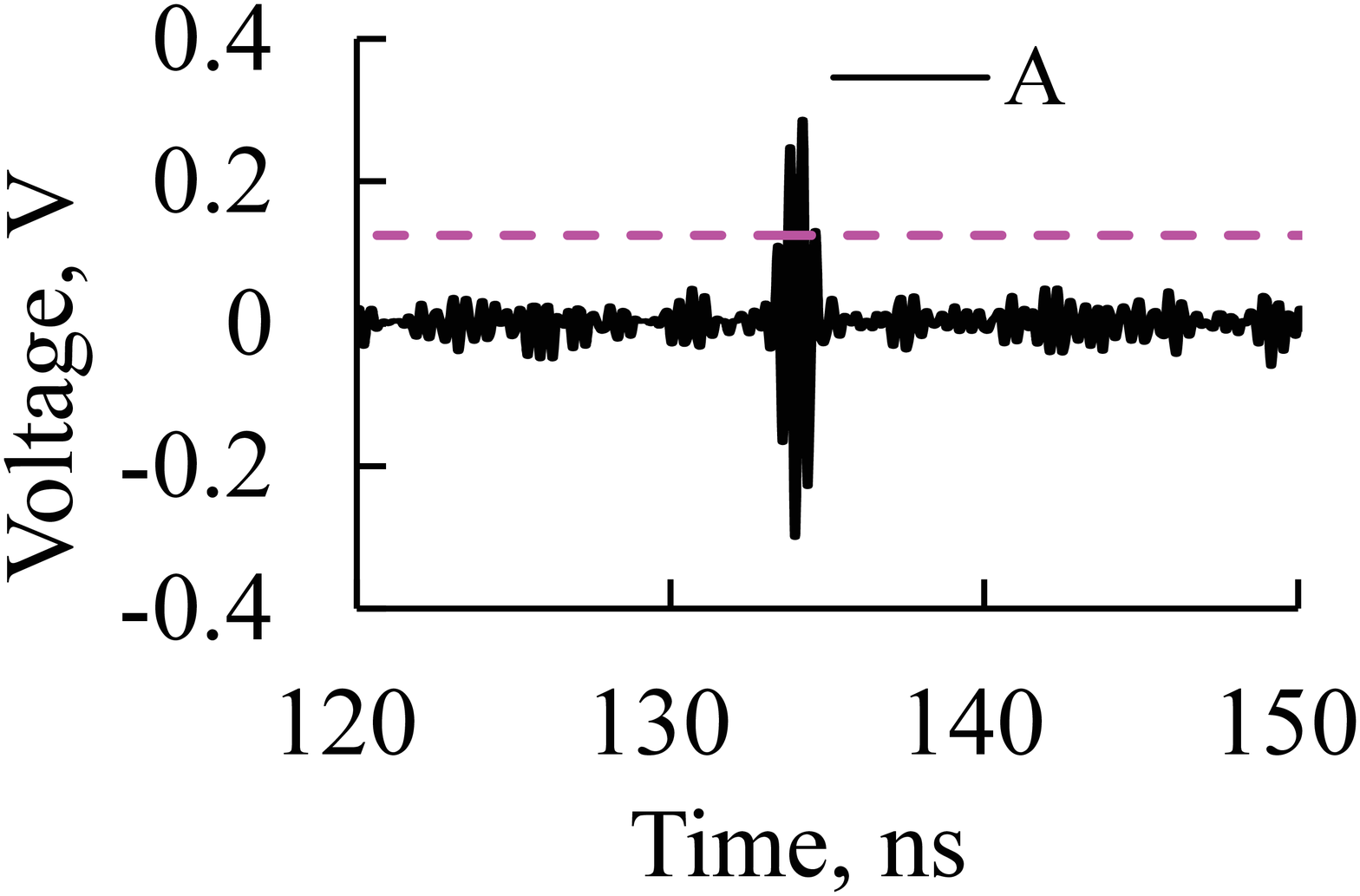}}
\caption{Signals received by the antennas (experiment). a) and b) denote the case when antenna-$1$ is set to be the target antenna. c) and d) denote the case when antenna-$2$ is set to be the target} \label{Fig:Case2}
\end{center}
\end{figure}

The same equivalent channel sounding method, based on channel reciprocity, as in~\cite{Ge-2011-subwavelegnth,Ding-2013-subwavelegnth}, is used in the calibration phase, where the sounding signals are actually generated by the transmitter and measured by the receivers, for simplicity. The experimental procedure is as follows:
\begin{enumerate}
  \item generate the channel sounding pulse, using the arbitrary wave generator, and transmit it with the time-reversal transmitter;
  \item record the signals received by the receiving antennas using the digital serial analyzer, flip them and numerically modulate them using a computer;
  \item transmit the modulated time-reversed signals from the time-reversal transmitter and record the signals received by both the target antenna and the non-target antenna.
\end{enumerate}
The experimental results are shown in Fig.~\ref{Fig:Case2}. Thanks to the PSCs, the waveform of the signal received by the target receiver is much higher and essentially identical to the transmitted one (not shown) while the signal received by the non-target receiver has a totally different waveform and is spread out in time with much lower temporal power density.
\begin{figure}
\begin{center}
\subfigure[]{\label{fig:com1}
\psfrag{A}[l][l][0.85]{Antenna-$1$}
\includegraphics[width=0.48\columnwidth]{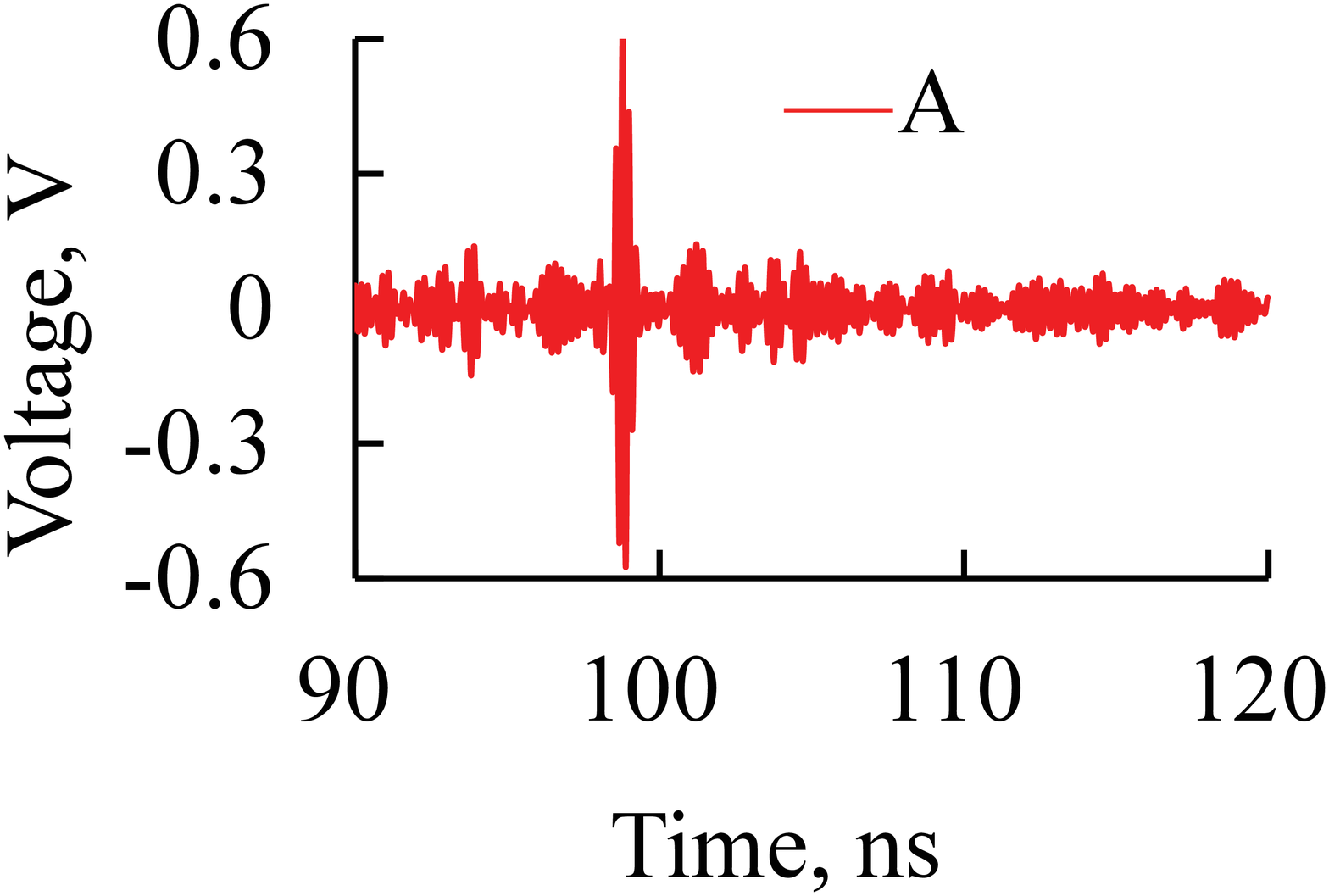}}
\subfigure[]{\label{fig:com2}
\psfrag{A}[l][l][0.85]{Antenna-$2$}
\includegraphics[width=0.48\columnwidth]{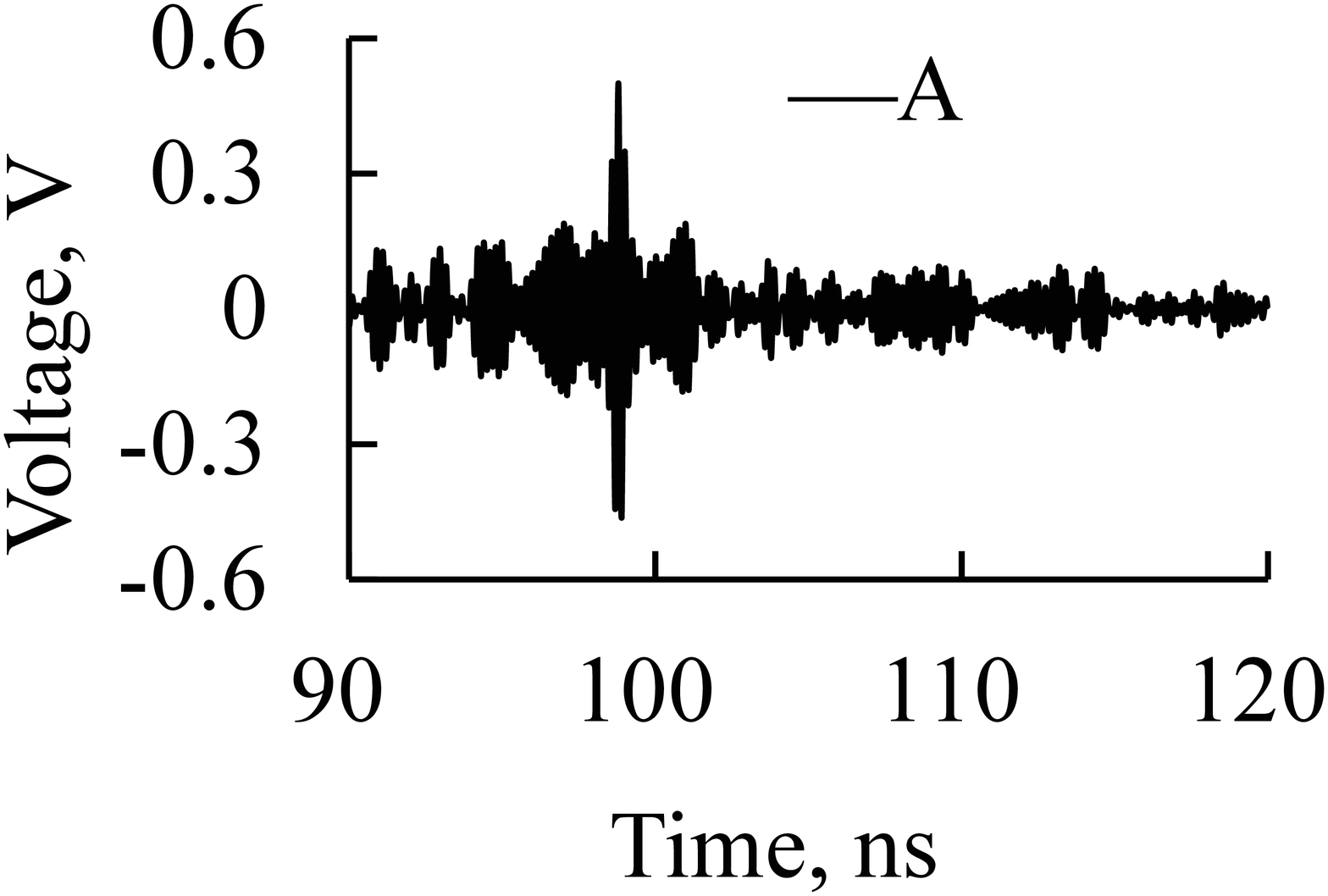}}
\caption{Same experiment as in Fig.~\ref{fig:a11} and ~\ref{fig:a21} but without pulse shaping circuits. a) Signal received by Antenna-1. b) Signal received by Antenna-2.} \label{Fig:compare}
\end{center}
\end{figure}

For comparison, the same experiment is carried out without PSCs. The results are shown in Fig.~\ref{Fig:compare}. In this case, as expected, the waveform of the signal received by the target receiver has lost its target features and is comparable to that received by the non-target receiver.

\section{Conclusions}\label{Sec:conclusion}

A novel approach to enhance time-reversal subwavelength transmission based on R-ASP PSCs is proposed, theoretically derived and experimentally validated. After a mathematical demonstration of the system, culminating with the determination of PSC conditions, experimental results are provided for validation, using two PSCs with different impulse responses to encode the channels between the transmitter and two receivers with sub-wavelength separation. The signals received by the antennas exhibit clear discrimination. Moreover, to the inherent analog nature of the PSCs, the time-reversal sub-wavelength transmission system is highly efficient and flexible. It may find applications in compact MIMO and higher resolution imaging systems.

\bibliographystyle{IEEEtran}
\bibliography{ReferenceList}

\end{document}